# Change-Point Detection Utilizing Normalized Entropy as a Fundamental Metric

1st Qingqing Song
Faculty of Applied Mathematics and Computer Science
Belarusian State University
Minsk, Belarus
fpm.sunC@bsu.by

2nd Shaoliang Xia
Faculty of Applied Mathematics and Computer Science
Belarusian State University
Minsk, Belarus
fpm.sya@bsu.by

*Abstract* – This paper introduces a concept for change-point detection based on normalized entropy as a fundamental metric, aiming to overcome the dependence of traditional entropy methods on assumptions about data distribution and absolute scales. Normalized entropy maps entropy values to the [0,1] interval through standardization, accurately capturing relative changes in data complexity. By utilizing a sliding window to compute normalized entropy, this approach transforms the challenge of detecting change points in complex time series—arising from variations in scale, distribution, and diversity—into the task of identifying significant features within the normalized entropy sequence, thereby avoiding interference from parametric assumptions and effectively highlighting distributional shifts. Experimental results show that normalized entropy exhibits significant numerical fluctuation characteristics and patterns near change points across various distributions and parameter combinations. The average deviation between fluctuation moments and actual change points is only 2.4% of the sliding window size, demonstrating strong adaptability. This paper provides theoretical support for change-point detection in complex data environments and lays a methodological foundation for precise and automated detection based on normalized entropy as a fundamental metric.

## I. Introduction

Change-point detection is a critical data analysis technique used to identify moments of significant statistical changes in time series or high-dimensional data [1], with broad applications in signal processing, financial analysis, industrial monitoring, and load analysis [2,3]. Common methods include statistical tests, model-based approaches, and information-theoretic techniques.

This paper focuses on entropy-based metrics within information theory for change-point detection. Entropy, introduced by Claude Shannon, measures the uncertainty or information content of a discrete random variable [4]. Recent advancements in this field include Sofronov et al. (2012), who applied cross-entropy to sequence change-point problems [5], and Unakafov and Keller (2018), who proposed a method based on conditional entropy of sequence patterns, effective in detecting structural changes but limited in pure level shifts [6]. Ma (2024) developed a non-parametric multivariate change-point detection method using copula entropy, achieving single and multiple change-point detection through two-sample tests and a binary segmentation strategy, with strong performance on simulated and real-world data [7].

This paper aims to introduce normalized entropy theory, design specific computational methods, and preliminarily validate the feasibility of using normalized entropy as a fundamental metric for change-point detection, while testing its applicability to multi-change-point detection on both simulated and real datasets.

## II. Utilizing Normalized Entropy as a Fundamental Metric

### A. Problem Statement

One of the core challenges in change-point detection lies in its reliance on assumptions about data distribution and specific numerical parameters. Many existing methods require pre-defined thresholds $\tau$, significance levels $\alpha$, or assumptions that data follow specific distributions $P(x)$, such as Gaussian or Poisson distributions [8, 9]. This dependency limits adaptability, as real-world time series often exhibit complex and diverse distributions that cannot be fully captured by a single assumption. Moreover, improper parameter settings can lead to biased results or reduced robustness.

For entropy-based change-point detection, the computation of entropy $H(X)$ typically assumes that the data distribution is known or can be estimated. However, in high-dimensional or nonlinear data, such assumptions may fail, causing entropy values to deviate from the true complexity. Furthermore, entropy changes are often defined as follows:

$$\Delta H = H(X_{t+1}) - H(X_t) \qquad (1)$$

The threshold setting $\Delta H > \tau$ often directly determines the outcome of change-point detection. If the threshold is set too high, it may lead to missed detections, while overly low thresholds can result in excessive false positives. These issues highlight that, despite the theoretical advantages of entropy in change-point detection, its practical application remains constrained by parameter selection and distributional assumptions. Designing robust, distribution-free change-point detection methods remains a key research challenge in this field.

*B. Eliminating Dependence on Absolute Entropy Scale*

The normalized entropy method addresses the dependence of traditional entropy metrics on absolute entropy scales, significantly improving adaptability and robustness in change-point detection. In classical entropy calculations, the value of $H(X)$ is directly influenced by the range, dimensionality, and scale of data. For datasets of varying sizes or time series with different dynamic ranges, entropy values may become difficult to compare due to absolute differences in data characteristics, making change-point detection results dependent on specific entropy scales.

Normalized entropy introduces a standardization factor to map entropy values to a unified range (typically [0,1]), providing a more accurate representation of the relative complexity or uncertainty of data. Formally, normalized entropy can be expressed as:

$$H_{norm}(X) = \frac{H(X)}{H_{max}} \quad (2)$$

Here, $H(X)$ represents the original entropy value, and $H_{max}$ denotes the maximum possible entropy, typically determined by the theoretical maximum complexity of the data. This normalization ensures that entropy changes are no longer dependent on absolute data scales but are instead driven solely by variations in data patterns, making it more suitable for diverse and non-uniform data environments.

*C. Approach and Algorithms*

To convert continuous data into discrete probability distributions, binning is required. Given a sample dataset $X = \{x_1, x_2, \cdots, x_n\}$, where $n$ is the sample size and the data range is $[x_{max}, x_{min}]$, the data is divided into $k$ equal-width intervals, with each interval width calculated as:

$$\Delta = \frac{x_{max} - x_{min}}{k} \quad (3)$$

The interval boundaries are:

$$\text{BinEdges} = \{x_{min}, x_{min} + \Delta, x_{min} + 2\Delta, \cdots, x_{max}\} \quad (4)$$

For each interval $I_j = [x_{min} + (j-1)\Delta, \ x_{min} + j\Delta)$ ($j = 1, 2, \cdots, k-1$, with the last interval as the closed interval $[x_{min} + (k-1)\Delta, \ x_{max}]$), the number of data points X falling into each interval $f_j$ is counted.

$$f_j = |\{x_i \in X | x_i \in I_j\}| \quad (5)$$

Here, $|\cdot|$ denotes the cardinality of the set (i.e., the number of data points). The frequency $f_j$ for each interval is then converted into probability $p_j$:

$$p_j = \frac{f_j}{n}, j = 1, 2, \cdots, k \quad (6)$$

The probability distribution satisfies the normalization condition: $\sum_{j=1}^{k} p_j = 1$.

Based on the probability distribution $\{p_1, p_2, \cdots, p_k\}$, the entropy H(X) can be calculated using the formula:

$$H(X) = -\sum_{j=1}^{n} p_j \log_b p_j \quad (7)$$

The maximum entropy $H_{max}$ corresponds to a uniform distribution, where all interval probabilities are equal, $p_j = \frac{1}{k}, j = 1, 2, \cdots, k$. Substituting the uniform distribution into the entropy formula yields the maximum entropy:

$$H_{max} = \log_b k \quad (8)$$

Normalized entropy is defined by standardizing H(X) to the range [0, 1] as:

$$H_{norm}(X) = \frac{H(X)}{H_{max}} = \frac{-\sum_{j=1}^{k} p_j \log_b p_j}{\log_b k} \quad (9)$$

To ensure normalized entropy captures local rather than global characteristics, a sliding window mechanism is introduced. With a window size of $\delta$ and a step size of 1, the input data sequence $X = \{x_1, x_2, \cdots, x_T\}$, where $T$ is the total number of data points, is processed. For calculating normalized entropy at $x_T$, the data within the window is:

$$X_t = \{x_{t-\delta+1}, x_{t-\delta+2}, \cdots, x_t\}, t \geq \delta \quad (10)$$

For each window $X_t$, its normalized entropy $H_{norm}(X_t) \in [0, 1]$ is computed.

All normalized entropy values are recorded as the sequence $H$.

$$H = \{H_{norm}(X_\delta), H_{norm}(X_{\delta+1}), \cdots, H_{norm}(X_T)\} \quad (11)$$

III. EXPERIMENTS AND VALIDATION

This paper uses disk read-write monitoring sequences from a web server as an example of performance time series data to visualize the correlation between normalized entropy and the original data. (Algorithm parameters: $\delta = 70, k = \ln \delta$)

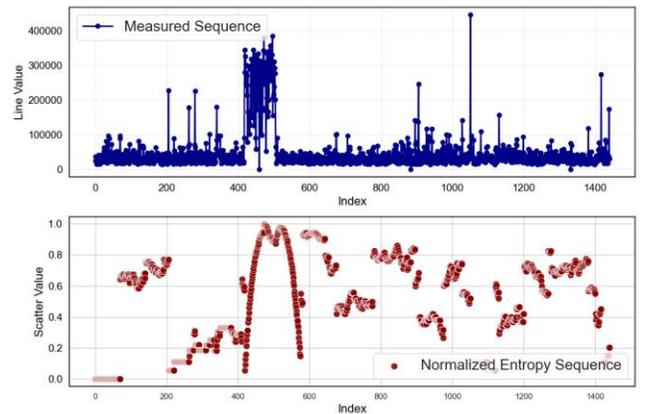

Fig. 1. Normalized Entropy and Original Sequence Comparison on Measured Data

As shown in Fig. 1, the comparison between the normalized entropy sequence and the original sequence clearly reveals significant local fluctuations in normalized entropy near change points, reflecting abrupt shifts in data complexity or uncertainty.

To further validate the clarity and stability of normalized entropy in characterizing change points, this paper conducts experiments using simulated data, generating time series with random combinations of varying lengths, parameters, and distributions as input. (Algorithm parameters: $\delta = 100, k = \ln \delta$)

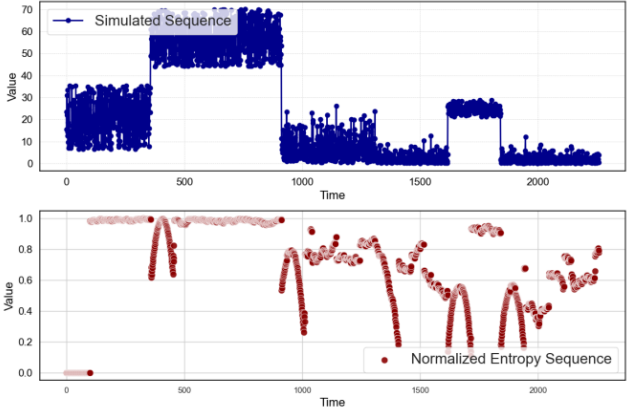

Fig. 2. Normalized Entropy and Original Sequence Comparison on Simulated Data

The validation results are shown in Fig. 2, which presents a time series composed of six combinations with varying levels of differences and distributions, along with their corresponding normalized entropy sequences. The original data exhibits significant shifts in numerical levels at five change points, while the normalized entropy sequence shows distinct fluctuations near these points, accurately aligning with the timing of abrupt changes in data distribution, as demonstrated in Table 1.

TABLE 1. CHANGE POINT AND NORMALIZED ENTROPY RESPONSE CHARACTERISTICS WITH ERROR ANALYSIS

| Change point (time) | Normalized Entropy Response Point (time) | Response Duration | Error |
|---|---|---|---|
| 357 | 359 | 98 | 2 |
| 912 | 914 | 98 | 2 |
| 1312 | 1316 | 93 | 4 |
| 1617 | 1619 | 98 | 2 |
| 1841 | 1843 | 98 | 2 |

According to Table 1, the average deviation between the fluctuation moments of the normalized entropy sequence and the actual change points is only 2.4% of the sliding window size, with the response duration close to the window size.

Near change points, the normalized entropy sequence typically forms prominent peaks or valleys, which can be characterized by local extrema, such as defining the peak of normalized entropy as:

$$H_{peak} = \max_{t \in [t-\delta, t+\delta]} H_{norm}(X_t) \qquad (12)$$

Here, $\delta$ represents the window size around the change point. These peak or valley features further validate the effectiveness of normalized entropy in change-point detection, accurately reflecting the locations of abrupt distribution changes and providing theoretical support for precise change-point localization.

## IV. DISCUSSION AND FUTURE RESEARCH DIRECTIONS

While the application of normalized entropy in change-point detection demonstrates theoretical advantages, its practical computation involves significant overhead. This paper employs a sliding window approach with a step size of 1, requiring the recalculation of normalized entropy for each time point, which increases computational complexity. Additionally, the choice of window size can impact both efficiency and detection accuracy. Optimizing the window mechanism or improving the algorithm to reduce computational costs is a key challenge for practical applications.

Future research could focus on clarifying the specific relationship between normalized entropy and change points and designing algorithms for precise localization. This paper highlights the distinct fluctuations of normalized entropy near change points but does not propose a concrete detection process. Observations suggest that these fluctuations often exhibit unique features, potentially containing rich pattern information. Further studies could analyze the local statistical properties of normalized entropy sequences to extract change-point characteristics. Methods like wavelet transform [10] or Fourier transform [11] could be applied to perform frequency-domain analysis of normalized entropy sequences, capturing characteristic frequencies or energy distributions near change points. Combining the dynamic behavior of normalized entropy with mathematical models or machine learning algorithms to establish automated change-point detection mechanisms is another important direction. Moreover, exploring alternative metrics with lower computational complexity, such as the DUID metric discussed in TRLLD [12], to quantify the uniformity of sample distribution density, offers a promising research avenue. Integrating sequential analysis methods and their related variants in time series is expected to address computational cost and real-time issues while improving the accuracy and efficiency of change-point detection [13].

## V. Conclusion

This paper proposes using normalized entropy as a fundamental metric for change-point detection to overcome the dependence of traditional entropy methods on data distribution assumptions and absolute scales. By standardizing entropy values, normalized entropy more accurately reflects relative complexity changes in data patterns, eliminating scale effects and providing theoretical support for change-point detection in diverse data environments. Methodologically, normalized entropy is computed via a sliding window, mapping entropy values to the [0,1] range, and characterizing significant fluctuations near change points through local extrema. This approach avoids interference from parametric assumptions and highlights abrupt changes in data distributions. Experiments demonstrate that normalized entropy reliably captures change points across various distributions and parameter combinations, showing strong robustness and adaptability. Its distinct fluctuations near change points effectively represent abrupt shifts in data distributions, validating its effectiveness as a fundamental metric for change-point detection. Future research will focus on precise localization of change points, defining detection processes, and refining the practical theoretical framework based on normalized entropy.

## Acknowledgment

The authors wish to express their sincere appreciation to Professor Kharin Aleksey Y. for sharing his insights on change-point detection theory during his lecture, which greatly inspired this work.